\documentclass[3p,times,procedia]{elsarticle}
\usepackage{nupha_ecrc}

\volume{00}

\firstpage{1}

\journalname{Nuclear Physics A}

\runauth{N. R. Sahoo for the STAR Collaboration}

\jid{nupha}

\jnltitlelogo{Nuclear Physics A}

\usepackage{amssymb}

  \usepackage{lineno}

\usepackage[figuresright]{rotating}

\newcommand{\pT}{$\rm p_{T}$}

\newcommand{\sNN}{$\sqrt{{s}_{NN}}$}

\newcommand{\DirPho}{$\gamma_{dir}$}
\newcommand{\piZro}{$\pi^{0}$}
\newcommand{\GammaRich}{$\gamma_{rich}$}

\begin{document}
\begin{frontmatter}




\title{Jet Measurements with Neutral and Di-jet Triggers in Central Au+Au Collisions at \sNN~ = 200 GeV with STAR}


\author{Nihar Ranjan Sahoo (for the STAR Collaboration)}

\address{Cyclotron Institute, Texas A$\&$M University, USA\\ Email id: nihar@rcf.rhic.bnl.gov}

\begin{abstract}
 
We present two measurements related to jet production in p+p and central Au+Au collisions at \sNN=200 GeV. Firstly, a study of semi-inclusive charged recoil jets coincident with high-\pT~direct-photon and neutral pions, and secondly, the hadron correlations with respect to reconstructed di-jet triggers. Indication of medium effects is observed by comparing $\gamma$+jet and \piZro+jet measurements. The di-jet+hadron study shows signs that the medium-induced modifications of an imbalanced set of di-jets with ``hard cores'' primarily affect the recoil jet.

\end{abstract}


\end{frontmatter}


\section{Introduction}
\label{S1}
Jets and their modifications due to partonic energy loss provide a powerful tool to study the properties of the Quark Gluon Plasm (QGP) created in ultra-relativistic heavy-ion collisions. 
In order to study parton energy loss, we exploit the fact that different measurements have different trigger biases and explore the effect of the medium on jets for different types of triggers: i) \DirPho~{\it vs.} \piZro~and ii) di-jet triggers. The \DirPho~triggers do not suffer from a surface bias and carry approximately the initial energy of the recoil parton~\cite{Wang_Huang_Sarcevic,Wang_Zhu}; while comparing \DirPho~and \piZro~triggers offer the additional benefit of probing the color-factor dependence of the recoil parton energy loss. 

Recent $A_{J}$ measurements at STAR~\cite{Adamczyk:2016fqm} found a sample of di-jets selected with ``hard cores", i.e. only using constituents with \pT~$>$ 2 GeV/c,
that was imbalanced in central Au+Au collisions compared to a p+p reference. This lost energy was recovered in soft constituents (\pT~$>$
0.2 GeV/c) with signs of broadening of the jet structure within a jet radius of $R=0.4$. Charged hadron correlations with these di-jets enable us to investigate the redistribution of energy within the medium in more detail.  

 The STAR detector system provides full 2$\pi$ azimuthal coverage and pseudorapidity range within $|\eta|<1.0$. The Time Projection Chamber
(TPC) is used as the charged-particle tracking device~\cite{STAR_TPC}. The Barrel Electromagnetic Calorimeter (BEMC) ~\cite{STAR_BEMC} is used for
triggering and measuring the neutral trigger (photon or \piZro) and the neutral energy of a jet.  Events were selected by an online high tower (HT) trigger, which required an uncorrected transverse tower energy of $E_{T} >$ 5.4 GeV in at least one BEMC tower. 
 \vspace{-10pt}
\section{Semi-inclusive neutral-trigger recoil jets}
\label{S3}
 The charged recoil jets are reconstructed using the anti-$k_{T}$ algorithm from the FastJet package ~\cite{Cacciari:2011ma} with a jet resolution parameter of $R=0.3$, using charged tracks with \pT $>$ 0.2 GeV/c. The charged recoil jets are selected within the azimuthal angle between the neutral trigger and recoil jets satisfying $\Delta \phi \in [\frac{3\pi}{4},\frac{5\pi}{4}]$. The direct-photon hadron correlation analysis ~\cite{STAR_PLB_jetLike} provides a technique to trigger on direct photons using the transverse shower profile (TSP)
method which helps to discriminate between \GammaRich~(enriched sample of direct photon) and \piZro. 
The uncorrelated background jets are subtracted using a mixed-event (ME) method~\cite{STAR_hjet}. 
\begin{figure*}
\centering
 
  \includegraphics[width=0.4\linewidth]{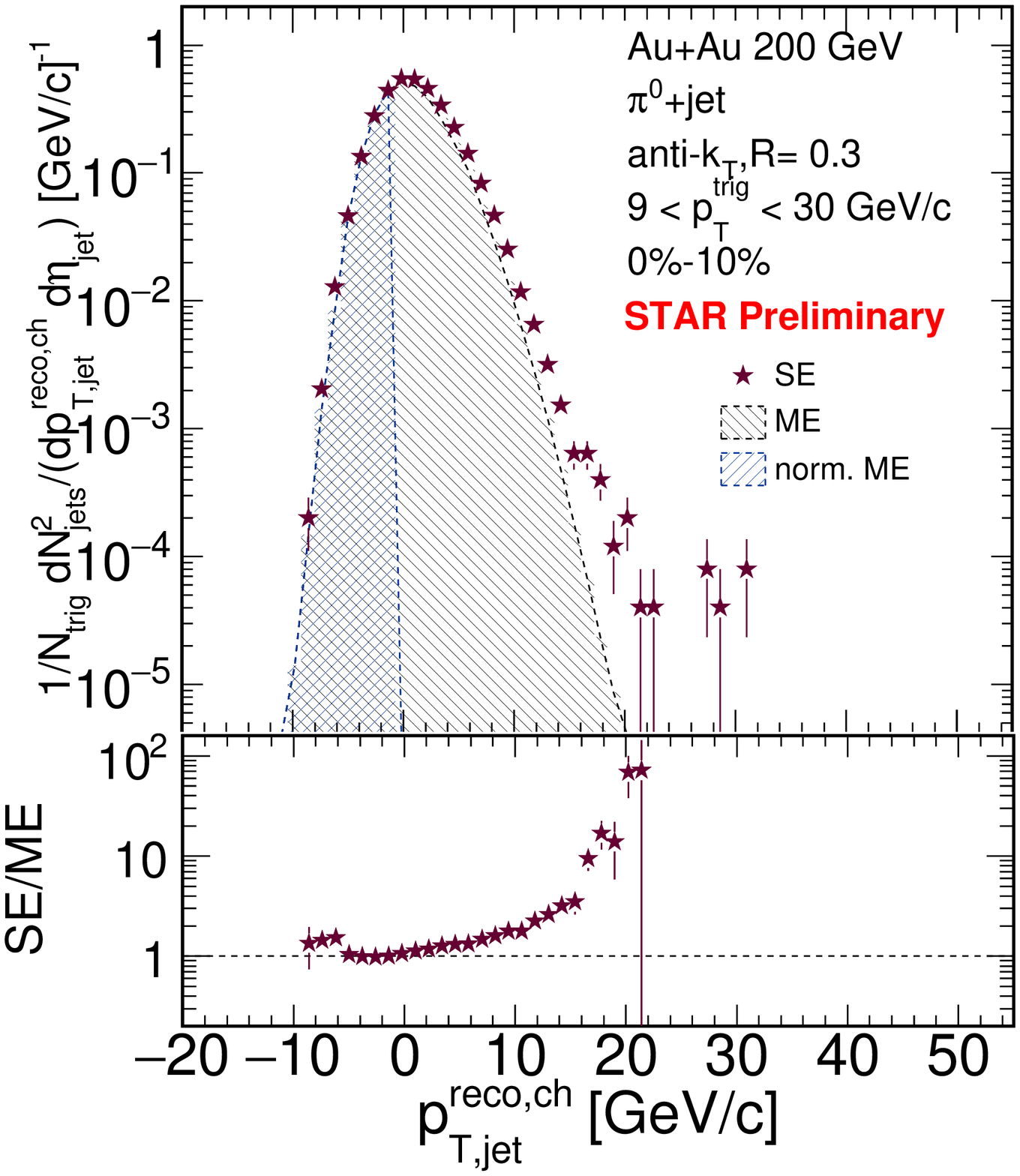}
   \includegraphics[width=0.4\linewidth]{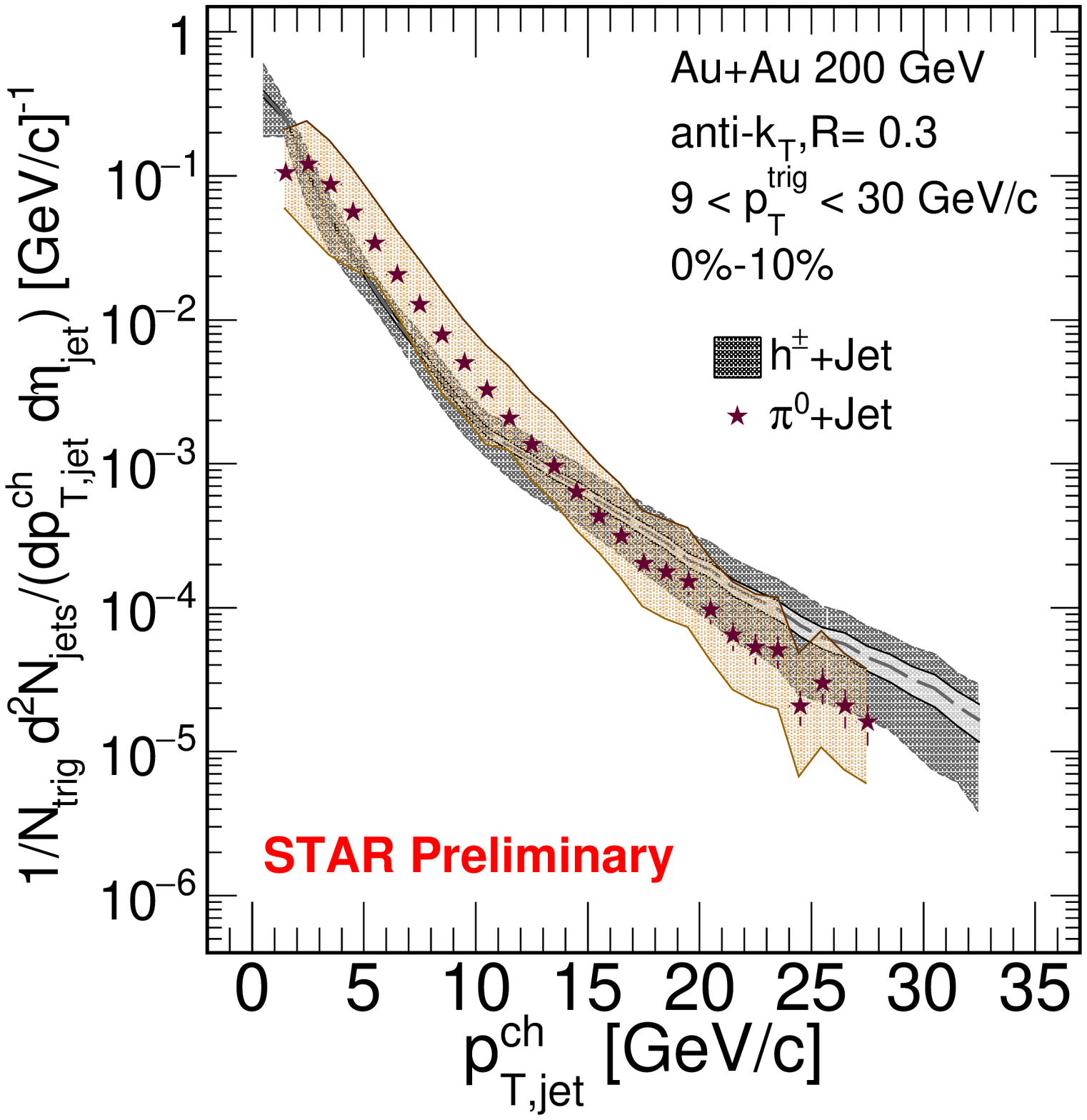}

 \vspace{-10pt}
 \caption{Left panel: The $\rm {p_{T,jet}^{reco,ch}}$ distribution of \piZro+jet in 0\%-10\% central Au+Au collisions at \sNN~=200 GeV. SE (star marker), ME (black shaded region) and norm. ME (blue shaded region) represent same event, mixed event, and the normalization region, respectively. Right panel: The corrected $\rm {p_{T,jet}^{ch}}$ distribution of both \piZro+jet and $\rm{h^{\pm}+jet}$~\cite{STAR_hjet} in 0\%-10\% central Au+Au collisions at \sNN~=200 GeV. The shaded bands represent the systematic uncertainty. 
 }
  \label{Fig1}
     \vspace{-10pt}
\end{figure*}

\begin{figure*}
\centering
   \includegraphics[width=0.34\linewidth]{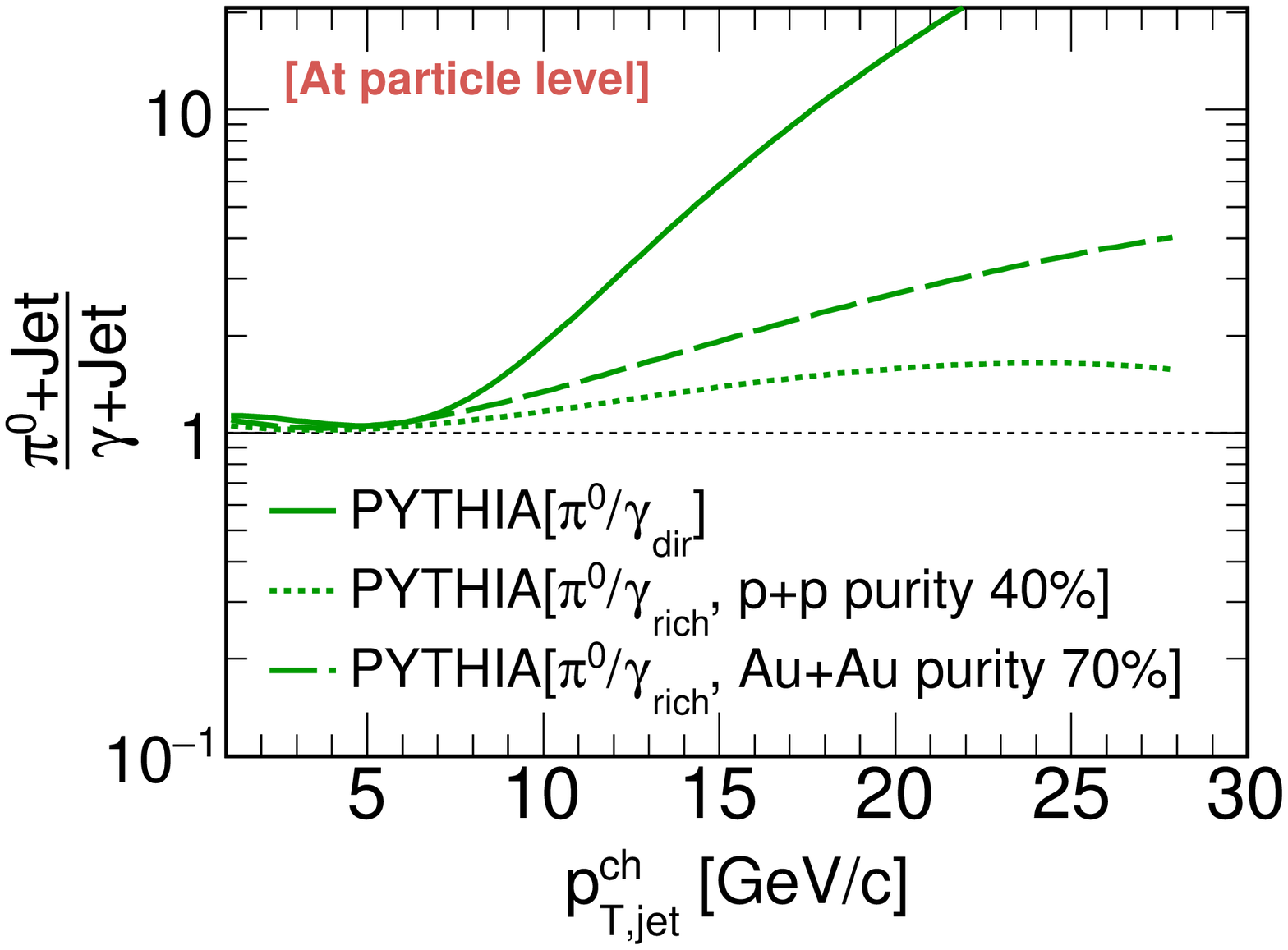}	
    \includegraphics[width=0.34\linewidth]{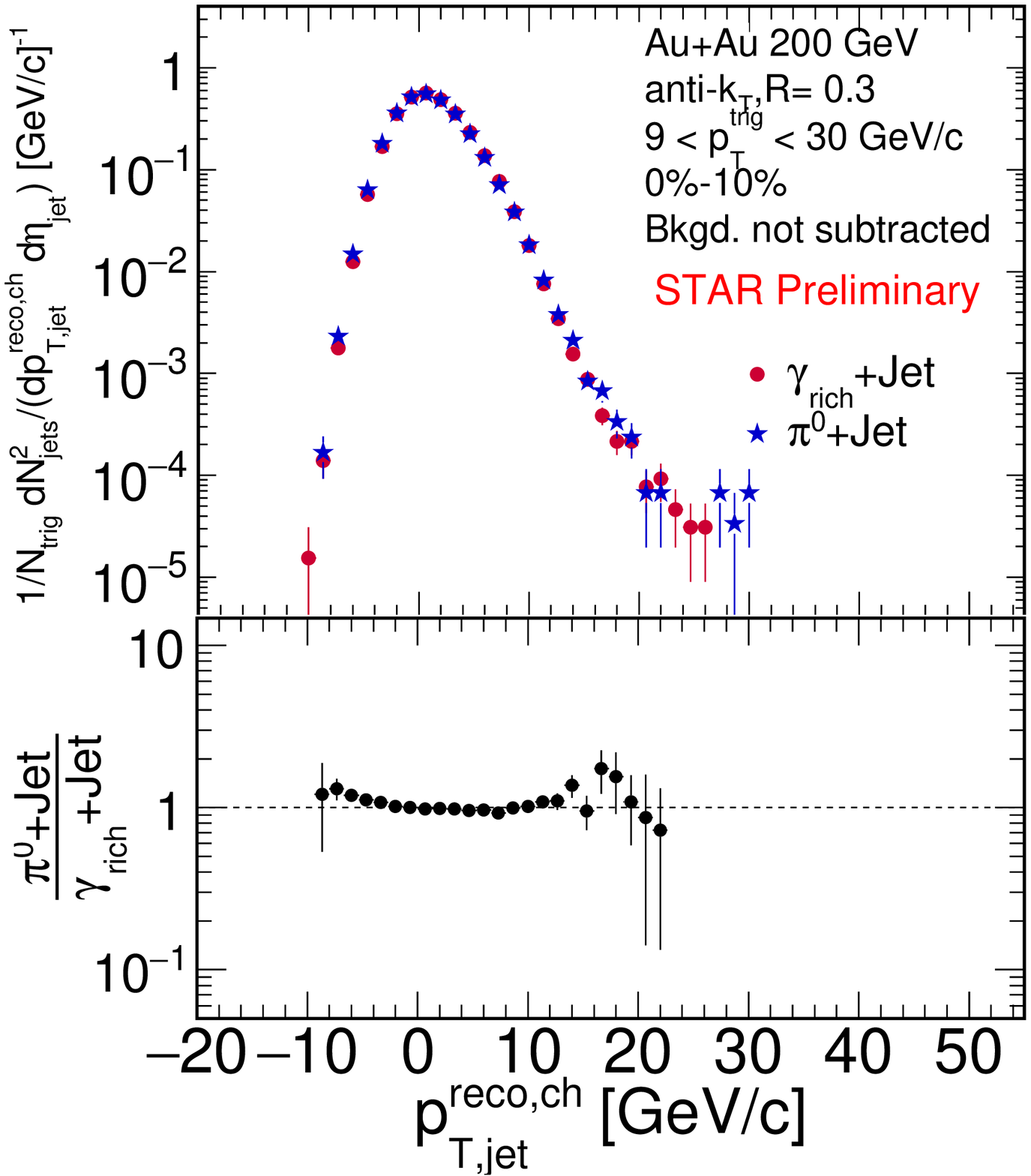}

\vspace{-10pt}
 \caption{
Left panel: Ratio
of recoil yields for $\pi^0$+jet and $\gamma$+jet as a function of $\rm {p_{T,jet}^{ch}}$ at particle level in the PYTHIA simulation. Solid, dotted and broken lines represent 100\%, 70\% (in Au+Au~\cite{STAR_PLB_jetLike}) and 40\% (in p+p~\cite{STAR_PLB_jetLike}) purity level in direct-photon sample in the PYTHIA simulation, respectively. 
 Right panel: The $\rm {p_{T,jet}^{reco,ch}}$ distributions of \piZro+jet (star marker) and \GammaRich+jet (circle marker) in 0\%-10\% central Au+Au collisions at \sNN~=200 GeV. 
 }
  \label{Fig2}
   \vspace{-5pt}
\end{figure*}

The distribution of reconstructed charged recoil jet transverse momentum ($\rm {p_{T,jet}^{reco,ch}}$) of \piZro~trigger is shown on the left side of Fig.~\ref{Fig1}. The raw reconstructed jet energy for each jet candidate $i$ is corrected for the estimated average background energy contained within the jet. Here, $\rm {p_{T,jet}^{reco,i}} = \rm{p_{T,jet}^{raw,i}} - \rho \cdot \rm{A^{i}_{jet}} $, where $\rm{p_{T,jet}^{raw,i}}$ is the \pT~of $i^{th}$ jet candidate measured by the jet reconstruction algorithm, $\rm {A^{i}_{jet}}$ is the area of the jet and the background energy density, $ \rho \equiv \rm{median[\frac{\rm{p_{T,jet}^{raw,i}}} {  {{A^{i}_{jet}}}}]}$ (using the $k_{T}$ algorithm). At high jet \pT, the correlated recoil signal dominates the background. The integral of the ME jet \pT~distribution is normalised within the purely combinatorial region with $\rm {p_{T,jet}^{reco,ch}} < 0$ to that of the same events (SE). The correlated jet \pT~distribution is obtained by subtracting the normalised ME from SE. This subtracted jet \pT~distribution is then corrected by an unfolding procedure (adopting the singular value decomposition method ~\cite{SVD}) for detector efficiency and \pT-smearing due to background effects.
 The right panel of Fig.~\ref{Fig1} shows a comparison between the reconstructed charged recoil jet spectra triggered by \piZro~and $\rm{h^{\pm}}$~\cite{STAR_hjet} after corrections.
  The spectra show agreement within systematic uncertainty. Additional measurements are ongoing to reduce the systematic and statistical uncertainty.

A comparison between \piZro+jet and $\gamma$+jet distributions is performed at the particle level PYTHIA~\cite{PYTHIA} simulation as shown in Fig.~\ref{Fig2}. In this simulation, the purities of direct photons and \piZro~are varied from the ideal to estimate the influence of impurity. 
When assuming 100\% direct-photon and \piZro~purity, it is found that the ratio of the jet yield recoiling from a \piZro~compared to a direct-photon increases as a function of $\rm{p_{T,jet}^{ch}}$ (above $\rm{p_{T,jet}^{ch}}$~$>$10 GeV/c): a trigger \piZro~carries on an average of 85\% of total energy of its parent parton~\cite{STAR_PLB_jetLike} in contrast to the direct-photon trigger. In our data, we estimate the direct-photon purity to be 70\% ~\cite{STAR_PLB_jetLike} in Au+Au collisions (40\% in p+p collisions) and hence we denote the trigger sample as ``\GammaRich". For lower \DirPho~purities, the enhancement is reduced (due to the contamination in the population) but still present. The measured yield ratio, however, is consistent with unity with a large statistical uncertainty in 0\%-10\% central Au+Au collisions as shown on the right side of Fig.~\ref{Fig2}. This difference could be due to the medium effect; larger statistics and complete corrections, such as detector efficiency and background fluctuations effects, are needed to draw a strong  conclusion.

\vspace{-12pt}
 \section{Di-jet+hadron correlations}
\label{S4}
 \vspace{-3pt}
The di-jet selection for this analysis is similar to that done in the STAR  $A_J$ analysis~\cite{Adamczyk:2016fqm}, requiring a pair of back-to-back ``hard core'' di-jets such that $\rm {p_T^{\mathrm{lead}}}>20.0$ GeV/$c$ and $\rm {p_T^{\mathrm{sublead}}}>10.0$ GeV/$c$.  The back-to-back jets are reconstructed with a hard constituent cut of $\rm{p_T^{\mathrm{const}}}>2.0$ GeV/$c$ and $R=0.4$ using the anti-$k_T$ algorithm from the FastJet package~\cite{Cacciari:2011ma}. Events were selected to have an online calorimeter deposition of $E_{T} >$ 6 GeV, and after hadronic correction, a tower with $E_{T} >$ 5.4 GeV is required in one of the reconstructed jets. Jets are distinguished between trigger and recoil jets depending on which jet contains the High Tower that fulfilled the trigger requirement. To match the jet energy scale between Au+Au and p+p, the lower tracking efficiency in central Au+Au events must be taken into account~\cite{Adamczyk:2016fqm}. The p+p jet energy scale was corrected by randomly discarding charged tracks before jet finding with a probability of 1 - ($\epsilon_{\mathrm{Au+Au}}(\rm{\eta,p_T})/\epsilon_{\mathrm{p+p}}(\rm{\eta,p_T})$), where $\epsilon$ is the measured TPC efficiency for that collision system. To account for background fluctuations, the p+p data are embedded into minimum bias, 0\%-20\% central Au+Au events for jet-finding.    

\begin{figure}
\begin{center}
\includegraphics[scale=0.2]{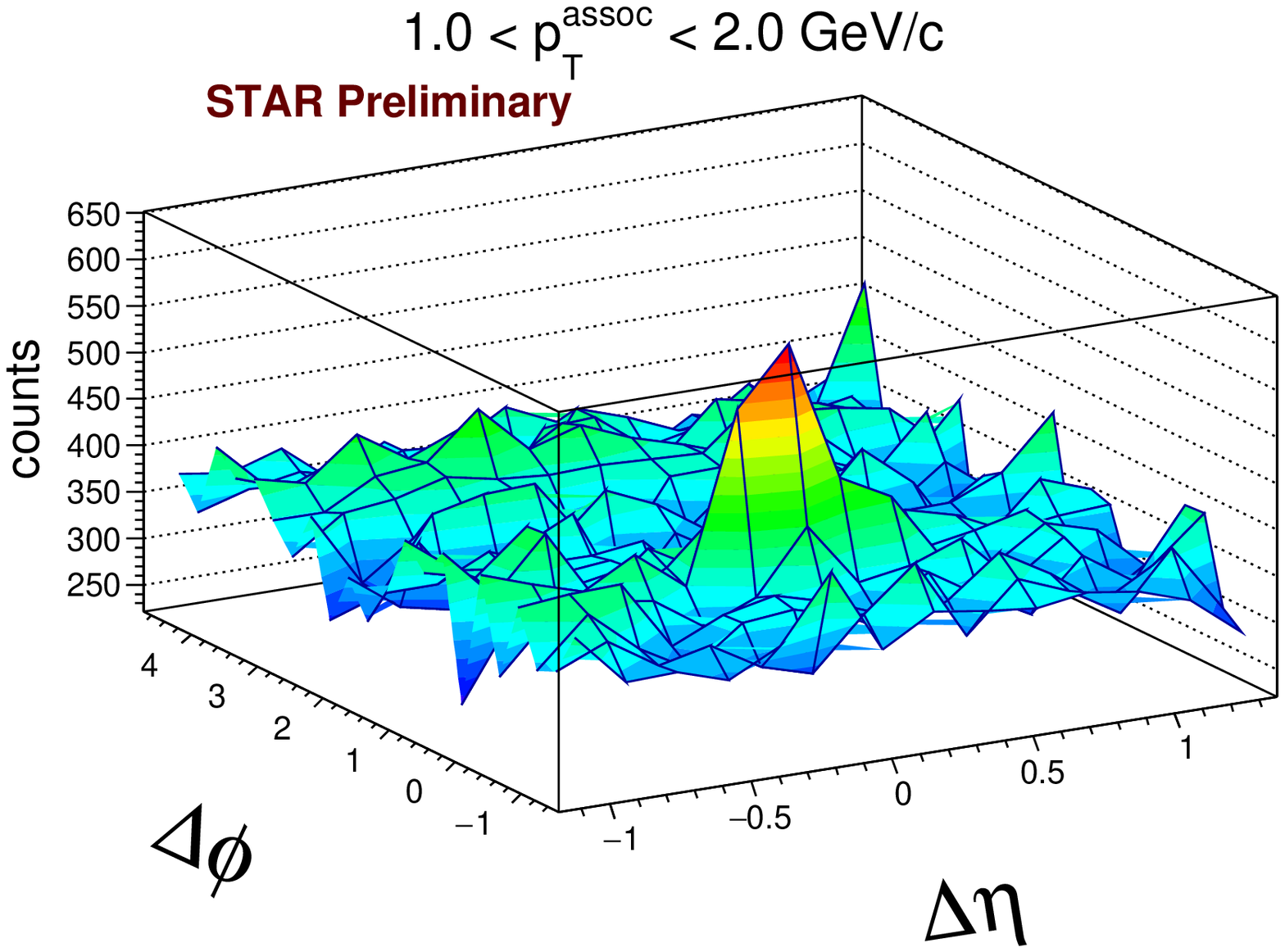}
\includegraphics[scale=0.25]{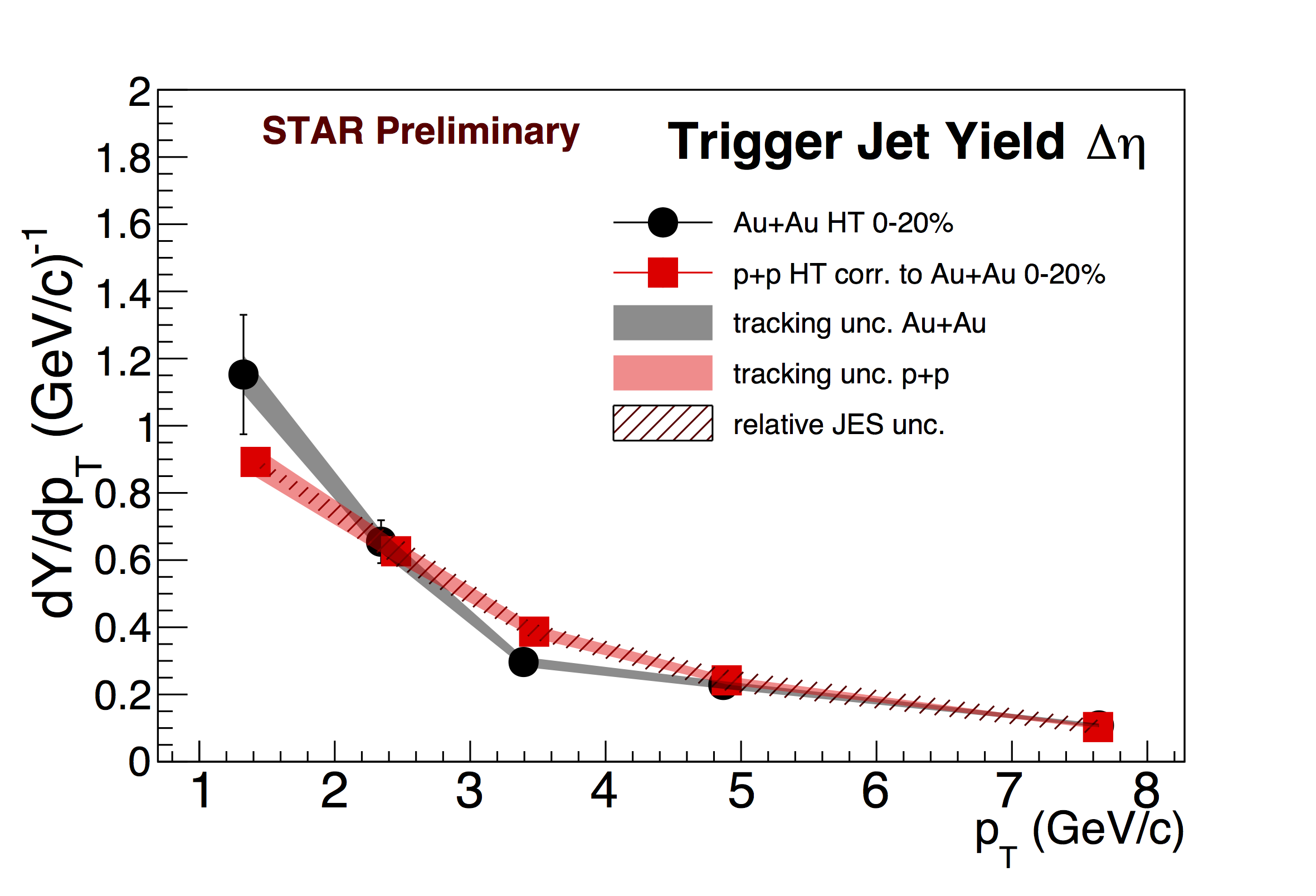}
\includegraphics[scale=0.25]{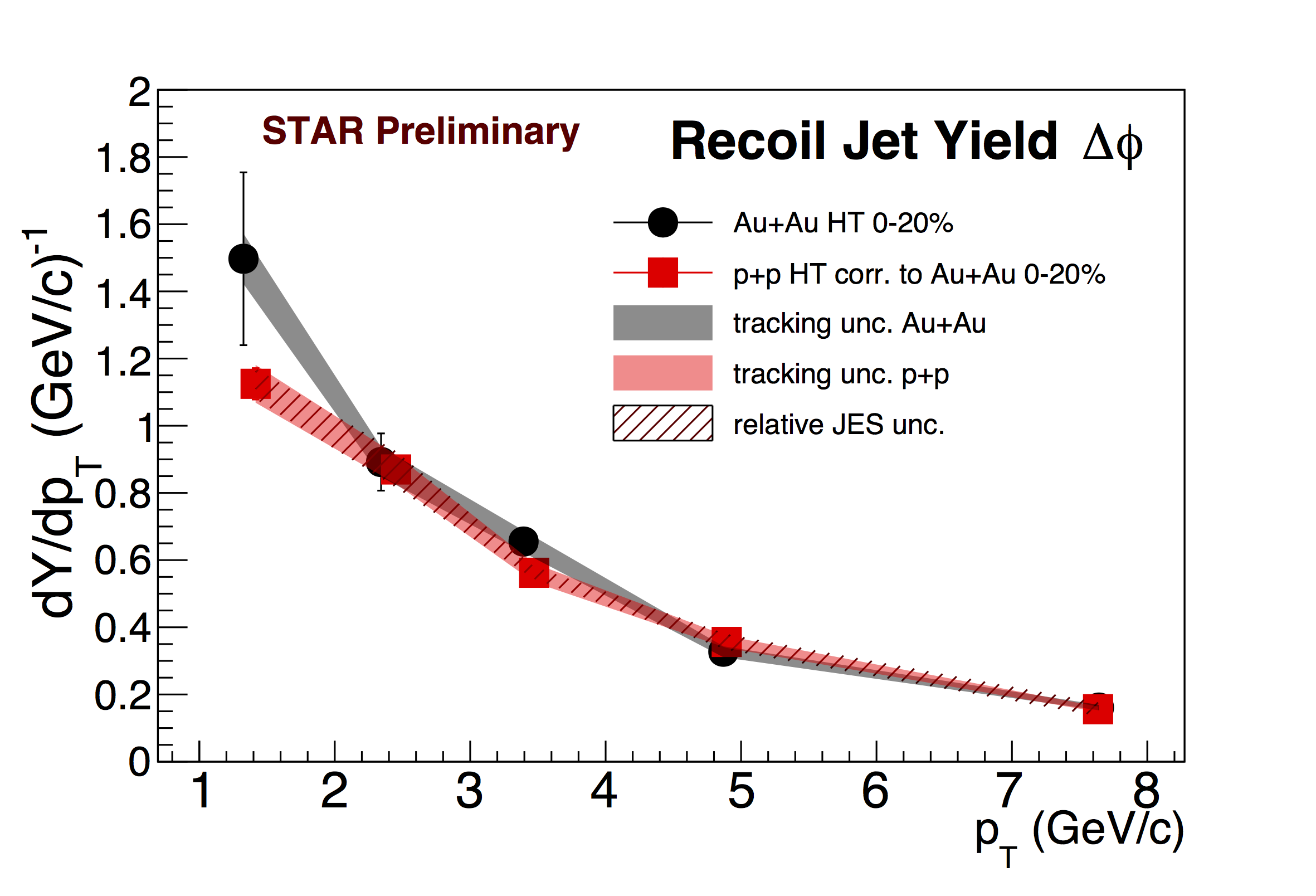}
\caption{
Left: 2D correlations after mixed event correction. 
Middle: Trigger jet yields in $\Delta\eta$ for Au+Au (circle) and p+p (box). Right: Recoil jet yields in $\Delta\phi$ for Au+Au (circle) and p+p (box). The bands represent the systematic uncertainty. }

  \label{dijets}

 \vspace{-23pt}
\end{center}
\end{figure}  
 
Correlations are measured for both trigger and recoil jets with all charged tracks, in both $\Delta \eta = \eta^{\mathrm{jet}}-\eta^{\mathrm{const}}$ and $\Delta \phi = \phi^{\mathrm{jet}}-\phi^{\mathrm{const}}$ as a function of $\rm {p_T^{\mathrm{const}}}$. Single particle tracking efficiency was corrected to particle level
using $\epsilon_{\mathrm{Au+Au}}(\rm {\eta,p_T})$ and $\epsilon_{\mathrm{p+p}}(\rm {\eta,p_T})$ in central Au+Au and p+p collisions, respectively. To correct pair-acceptance effects, the mixed event technique is used. The raw signal correlations are divided by the mixed event distribution in $\rm {p_T^{\mathrm{const}}}$ bins; an example can be seen in the left panel of Fig. 3. The distribution is flat out to large $\Delta \eta$ on the near side, has a strong near-side peak around the trigger jet axis, and an away side ridge around $\Delta \phi \approx \pi$.

Yields are calculated in $\Delta \eta$ and $\Delta \phi$ by subtracting the background and integrating over bins. $\Delta\eta$ background subtraction is performed by fitting with a gaussian + constant and subtracting the constant. $\Delta\phi$ background subtraction must account for long range $\Delta \eta$-independent correlations between the jet axis and the underlying event. It is assumed that any flow ($v_n$) correlations are independent of $\Delta \eta$. The side band subtraction method is used, where the region $0.45<|\Delta\eta|<1.0$, properly scaled, is used to subtract background and any potential correlation with the underlying event from the signal region defined by $|\Delta\eta|<0.45$. 

Yields extracted from $\Delta\phi$ and $\Delta\eta$ projections are consistent within uncertainty. The trigger jet yield in $\Delta\eta$ shows no statistically significant modification at any $\rm{p_T^{\mathrm{const}}}$ (middle panel of Fig.~\ref{dijets}), implying a strong surface bias, as expected from requiring a high energy trigger hadron. On the other hand, recoil jets show hint of increased correlated yield with respect to p+p at low $\rm{p_T^{\mathrm{const}}}$. Since the initial jet-finding is done with $\rm{p_T^{\mathrm{const}}}>2.0$ GeV/$c$, it is by construction that modification is observed mainly below the 2.0 GeV/$c$ threshold. This is also consistent with the interpretation of the $A_J$ results~\cite{Adamczyk:2016fqm}, where balance was restored when constituents below 2.0 GeV/$c$ were included. It is observed in the $A_J$ measurement that a large part of the di-jet population in Au+Au is balanced (p+p-like). As the correlations integrate over all $\Delta A_J=A_J^{\mathrm{hard}}-A_J^{\mathrm{full}}$, this can lead to a reduction in the yield differences between Au+Au and p+p.

 \vspace{-13pt}
\section{Summary and outlook}
\label{S5}
Comparison between \piZro+jet and $\rm{h^{\pm}+jet}$ distributions shows agreement within the systematic uncertainties. Larger statistics and final corrections are needed to draw strong conclusions about the potential medium effects when comparing $\gamma$+jet and \piZro+jet measurements in Au+Au collisions at RHIC.  The di-jet+hadron correlations study supports the picture that the selected di-jet sample with hard cores is mostly tangential to the fireball, and that while the signal is diluted by unmodified jets, modification primarily affects the recoil jet.  An increase in statistics from newer data sets (from Au+Au collisions measured by STAR in years 2014 and 2016) will allow for precise measurements of jets recoiling off a high-\pT~neutral particle, as well as enable differential studies of di-jet imbalance and
potential path length control via ``jet geometry engineering".






\vspace{-10pt}
\section*{Acknowledgments}
\vspace{-8pt}
  
This work was supported by the US DOE under the grant DE-FG02-07ER41485 and DE-FG02-92ER-40713.

 \vspace{-13pt}
\bibliographystyle{elsarticle-num}
\bibliography{<your-bib-database>}



\end{document}